\documentstyle[preprint,psfig,aps,epsf,floats]{revtex} 

\newcommand{\met}{\mbox{${\hbox{$E$\kern-0.6em\lower-.1ex\hbox{/}}}_T$}}
\newcommand{\ppbar}{$p\overline{p} $}            
\newcommand{\ipb}{pb$^{-1}$}                     
\newcommand{\gevcc}{\mbox{GeV/$c^2$}}      
\newcommand{\gevc}{\mbox{GeV/$c$}}
\def\wino#1{\mbox{$\widetilde \chi^{\pm}_{1}$}}  
\def\zino#1{\mbox{$\widetilde \chi^{0}_{1}$}}  
\def\squark{\mbox{$\widetilde q$}}             
\def\gluino{\mbox{$\widetilde g$}}             
\def\mzero{\mbox{$m_0$}}                    
\def\mhalf{\mbox{$m_{1/2}$}}                
\def\mgluino{\mbox{$m_{\widetilde g}$}}             
\def\msquark{\mbox{$m_{\widetilde q}$}}             
\def\mchargino{\mbox{$m_{\wino{1}}$}}
\def\tanb{\mbox{$\tan \beta$}}
\def\dpipi{\mbox{$D_{\pi \pi}$}}
\def\D0{D\O}                              
\def\detadphi{\mbox{$\Delta \eta \times \Delta \phi$}} 

\def\Ht{\mbox{$H_T$}}                          
\def\ttbar{\mbox{$t\overline{t}$}}

\tightenlines
\begin{document}

\preprint{FERMILAB-Pub-98/402-E}
\title{Search for Squarks and Gluinos in Events Containing Jets and a 
Large Imbalance in Transverse Energy}

%
\author{                                                                      
B.~Abbott,$^{40}$                                                             
M.~Abolins,$^{37}$                                                            
V.~Abramov,$^{15}$                                                            
B.S.~Acharya,$^{8}$                                                           
I.~Adam,$^{39}$                                                               
D.L.~Adams,$^{49}$                                                            
M.~Adams,$^{24}$                                                              
S.~Ahn,$^{23}$                                                                
G.A.~Alves,$^{2}$                                                             
N.~Amos,$^{36}$
% ...
\\                                                                            
\centerline{(D\O\ Collaboration)}                                             
}                                                                             
\address{                                                                     
\centerline{$^{1}$Universidad de Buenos Aires, Buenos Aires, Argentina}       
\centerline{$^{2}$LAFEX, Centro Brasileiro de Pesquisas F{\'\i}sicas,         
                  Rio de Janeiro, Brazil}                                     
\centerline{$^{3}$Universidade do Estado do Rio de Janeiro,                   
                  Rio de Janeiro, Brazil}                                     
\centerline{$^{4}$Universidad de los Andes, Bogot\'{a}, Colombia}             
\centerline{$^{5}$DAPNIA/Service de Physique des Particules, CEA, Saclay,     
                  France}                                                     
\centerline{$^{6}$Panjab University, Chandigarh, India}                       
\centerline{$^{7}$Delhi University, Delhi, India}                             
\centerline{$^{8}$Tata Institute of Fundamental Research, Mumbai, India}      
\centerline{$^{9}$Kyungsung University, Pusan, Korea}                         
\centerline{$^{10}$Seoul National University, Seoul, Korea}                   
\centerline{$^{11}$CINVESTAV, Mexico City, Mexico}                            
\centerline{$^{12}$Institute of Nuclear Physics, Krak\'ow, Poland}            
\centerline{$^{13}$Institute for Theoretical and Experimental Physics,        
                   Moscow, Russia}                                            
\centerline{$^{14}$Moscow State University, Moscow, Russia}                   
\centerline{$^{15}$Institute for High Energy Physics, Protvino, Russia}       
\centerline{$^{16}$University of Arizona, Tucson, Arizona 85721}              
\centerline{$^{17}$Lawrence Berkeley National Laboratory and University of    
                   California, Berkeley, California 94720}                    
\centerline{$^{18}$University of California, Davis, California 95616}         
\centerline{$^{19}$University of California, Irvine, California 92697}        
\centerline{$^{20}$University of California, Riverside, California 92521}     
\centerline{$^{21}$Florida State University, Tallahassee, Florida 32306}      
\centerline{$^{22}$University of Hawaii, Honolulu, Hawaii 96822}              
\centerline{$^{23}$Fermi National Accelerator Laboratory, Batavia,            
                   Illinois 60510}                                            
\centerline{$^{24}$University of Illinois at Chicago, Chicago,                
                   Illinois 60607}                                            
\centerline{$^{25}$Northern Illinois University, DeKalb, Illinois 60115}      
\centerline{$^{26}$Northwestern University, Evanston, Illinois 60208}         
\centerline{$^{27}$Indiana University, Bloomington, Indiana 47405}            
\centerline{$^{28}$University of Notre Dame, Notre Dame, Indiana 46556}       
\centerline{$^{29}$Purdue University, West Lafayette, Indiana 47907}          
\centerline{$^{30}$Iowa State University, Ames, Iowa 50011}                   
\centerline{$^{31}$University of Kansas, Lawrence, Kansas 66045}              
\centerline{$^{32}$Louisiana Tech University, Ruston, Louisiana 71272}        
\centerline{$^{33}$University of Maryland, College Park, Maryland 20742}      
\centerline{$^{34}$Boston University, Boston, Massachusetts 02215}            
\centerline{$^{35}$Northeastern University, Boston, Massachusetts 02115}      
\centerline{$^{36}$University of Michigan, Ann Arbor, Michigan 48109}         
\centerline{$^{37}$Michigan State University, East Lansing, Michigan 48824}   
\centerline{$^{38}$University of Nebraska, Lincoln, Nebraska 68588}           
\centerline{$^{39}$Columbia University, New York, New York 10027}             
\centerline{$^{40}$New York University, New York, New York 10003}             
\centerline{$^{41}$University of Rochester, Rochester, New York 14627}        
\centerline{$^{42}$State University of New York, Stony Brook,                 
                   New York 11794}                                            
\centerline{$^{43}$Brookhaven National Laboratory, Upton, New York 11973}     
\centerline{$^{44}$Langston University, Langston, Oklahoma 73050}             
\centerline{$^{45}$University of Oklahoma, Norman, Oklahoma 73019}            
\centerline{$^{46}$Brown University, Providence, Rhode Island 02912}          
\centerline{$^{47}$University of Texas, Arlington, Texas 76019}               
\centerline{$^{48}$Texas A\&M University, College Station, Texas 77843}       
\centerline{$^{49}$Rice University, Houston, Texas 77005}                     
}                                                                             

\date{\today}

\maketitle
\begin{abstract}
Using data corresponding to an integrated luminosity of 79~\ipb, \D0\
has searched for events containing multiple jets and large missing
transverse energy in \ppbar\ collisions at $\sqrt{s}=1.8$~TeV 
at the Fermilab Tevatron collider.  Observing no significant excess 
beyond what is expected from the standard model, we set limits on the masses 
of squarks and gluinos and on the model parameters \mzero\ and \mhalf, in the 
framework of the minimal low-energy supergravity models of supersymmetry. For 
$\tanb = 2$ and $A_0 = 0$, with $\mu < 0$, we exclude all models with 
$\msquark <250$~\gevcc. For models with equal squark and gluino masses, we
exclude $m < 260$~\gevcc.
\end{abstract}


Supersymmetry (SUSY)~\cite{SUSY} is a symmetry that relates fermions
and bosons, and can solve the hierarchy problem of the Higgs sector of
the standard-model (SM)~\cite{hierarchy}.  Minimal SUSY extensions of
the SM (MSSM) require partners (sparticles) for all standard model particles:
a scalar partner for each quark and lepton (called squarks and
sleptons), and a spin-half partner for each of the gauge bosons and Higgs
scalars, which form the gluinos and the mixed states called charginos
and neutralinos.  Such models also require the presence of two Higgs
doublets, and thus four Higgs particles.  
Each particle in a SUSY model has an internal quantum
number called $R$-parity. If $R$ is conserved, as is assumed in this
analysis, then sparticle states must be produced in pairs, and each
sparticle that decays must contain an odd number of sparticles in its
decay products.  Consequently, in this scenario, the lightest SUSY
particle (LSP) must be stable, and can thereby provide a candidate for
dark matter.

The most general supersymmetric extension of the SM has over 100 
undetermined parameters.  Consequently, models
have been developed that contain additional symmetries and
constraints. Some of these involve Grand Unified Theories that include
a supersymmetry (SUSY-GUTs).  In this work, we consider the class of
models containing gravity-mediated SUSY breaking, called
supergravity (SUGRA) models~\cite{sugra}. In minimal low-energy
supergravity (MLES), the scalar (squark and slepton) masses are
unified to a single value \mzero\ at the GUT energy scale, and the
gaugino masses are unified to a single value \mhalf. Three other
parameters describe the Higgs sector of the model: \tanb, the ratio of
the vacuum expectation values of the two Higgs doublets; $A_0$, a
universal trilinear coupling constant; and the sign of $\mu$, a mixing
parameter in the Higgsino mass matrix. Models in which the
lightest neutralino (\zino{1}) is the LSP, the LSP interacts only
weakly. It therefore cannot be observed directly, providing an
excellent experimental SUSY signature: large missing transverse energy (\met). In
such models, squarks (\squark) and gluinos (\gluino) can decay through
a cascade of charginos and neutralinos to final states consisting of
quarks, leptons, and the LSP. 
In this Letter we describe a search for squarks and gluinos  
in the jets and \met\ channel. 

The data, corresponding to an integrated luminosity of $79.2\pm
4.2$~\ipb, were collected with the \D0\ detector\cite{detector} at the
Fermilab Tevatron \ppbar\ collider operating at a center-of-mass
energy of 1.8~TeV during 1993--95. \D0\ has three major
components: a central tracking system, central and forward
uranium/liquid-argon calorimeters with towers in pseudorapidity and
azimuth of $\detadphi = 0.1 \times 0.1$, and a toroidal muon
spectrometer. Jets are reconstructed using a cone
algorithm~\cite{jetshape} with a cone radius of 0.5 in $\eta-\phi$
space. The electromagnetic (EM) energy scale is set using the $Z
\rightarrow ee$ signal.  The jet energy scale is determined from
energy balance in events containing a hadronic jet and a photon
candidate. \met\ is calculated from the vector sum of energy deposited
in all calorimeter cells.

Events were collected using a trigger that required $\met > 40$~GeV, and
at least one calorimeter trigger tower (of size $\detadphi = 0.2
\times 0.2$) with transverse energy $E_T > 5$~GeV. Offline filtering
required $\met > 40$~GeV and at least two jets with $E_T > 8$~GeV.

Backgrounds arise from several sources, among which are \ttbar\
production and the production of $W$ and $Z$ bosons in association with
jets. Purely instrumental sources of background include QCD multijet production
in which jets are mismeasured, resulting in apparent \met.

To remove events with false large \met, due to detector noise and
losses from the accelerator, we required events to have a summed scalar 
$E_T$ ($S_T$) $0.0 < S_T < 1.8$~TeV.  This requirement has little effect 
on the efficiency for hard-scattering events with additional overlapping 
soft \ppbar\ interactions, because these contribute little $E_T$ 
(typical $<S_T>$ for a squark-glunio event is 400~GeV).
The position of the primary interaction vertex is required to be within
60~cm of the detector center. The initial data sample 
contains 71{,}023 events.  To reject events with large
\met\ caused by poorly measured jets, we required all jets with
$E_T > 15$~GeV meet quality criteria based on cluster
shape\cite{lyon}, and that the three jets with highest $E_T$ be within
$|\eta| < 1.1$, or within $1.4 < |\eta| < 3.5$. 
The shape requirements included rejecting
events in which any jet deposited more than 90\% of its energy in the
EM portion of the calorimeter; this 
significantly reduced the $W \rightarrow e \nu$ background.

To select events consistent with being hadronic decays of squarks and
gluinos, we required at least three jets with $E_T > 25$~GeV, and 
$\met > 75$~GeV.  Our trigger was fully efficient under these conditions.
We also required the leading jet to have $E_T > 115$~GeV, leaving 
544 events.  To suppress
QCD multijet background we required the azimuthal difference
between the \met\ and a jet of $E_T > 25$~GeV be $\delta \phi > 0.1$, or 
$<(\pi - 0.1)$ radians. To reject events where a fluctuation of the
second leading jet masks a fluctuation of the leading jet, we also
required $(\delta\phi_{1}-\pi)^2 + {\delta \phi_{2}}^2 \geq (0.5)^2$, 
where 1(2) denotes the leading (second-leading) jet in $E_T$.  To reduce
the background from $W$ and $Z$ boson production in association with
jets, we required $\Ht > 100$~GeV, where \Ht\ is defined as the scalar
sum of the transverse energies of all but the leading jet. To remove
the remaining $W
\rightarrow \mu \nu$ + jets events, we rejected events containing
isolated muons with $p_T > 15$~\gevc. A total of 49~events satisfied all
the above requirements.

The average Tevatron luminosity was $\approx 9 \times
10^{30}$~cm$^{-2}$~s$^{-1}$, and peaked at about $2 \times
10^{31}$~cm$^{-2}$~s$^{-1}$. For the average value, there is a 75\%
probability of having an additional $p \bar{p}$ interaction accompanying
the hard scattering. These additional events contribute many charged
tracks, which can occasionally cause the soft collision to be chosen as
the primary interaction vertex, and cause a gross mismeasurement of the
\met. To remove events with large \met\ caused by the
misreconstruction of the interaction point, we required the charged
tracks associated with the central jet of highest $E_T$ be consistent
with emanating from the primary interaction vertex~\cite{lyon}. The
15~events passing this criterion formed our penultimate event sample.

The final selection criteria for each $(\mzero, \mhalf)$ point were
determined by choosing \Ht\ and \met\ thresholds that maximized the
$S/\delta B$ ratio, where $S$ is the expected number of SUSY events
and $\delta B$ is the combined systematic and statistical error on the
background predicted from the SM.  Table~\ref{table:optimization}
shows the thresholds used, the number of events expected from SM
sources, and the number of events observed in the data.

\begin{table}[h]
\caption{Optimized \met\ and \Ht\ thresholds for several regions of MLES
parameter space. The optimal thresholds were chosen for the specified
$\mzero$ and $\mhalf$ values that correspond to the listed gluino and
squark masses. The next-to-leading order cross sections and the total
efficiency for signal events, with their combined statistical and
systematic uncertainties, the total number of events expected from
backgrounds, with their statistical and systematic uncertainties, the
number of observed events, the probability for observing $N_{\rm obs}$
events or greater given the background prediction, and the 95\%
confidence level upper limit on the cross section are given in the
remaining columns.  Note that the entries in this table are strongly
correlated.}
\label{table:optimization}
\begin{tabular}{c c c c r c r@{\hspace{0.1em}$\pm$\hspace{0.1em}} r@{\hspace{0.1em}} c@{\hspace{0.1em}} l r r r}
$\met_{\rm thresh}$ & $H_{T_{\rm thresh}}$ & ($\mzero$, $\mhalf$) & 
($m_{\tilde{g}}$, $m_{\tilde{q}}$) & \multicolumn{1}{c}{$\sigma_{\rm sig}$} & 
$\epsilon$ & \multicolumn{4}{c}{$N_{\rm bck-pred}$} & $N_{\rm obs}$ &  
\multicolumn{1}{c}{$P_{\rm over}$} & \multicolumn{1}{c}{$\sigma_{95}$}\\
(GeV) & (GeV) & (\gevcc) & (\gevcc) & (pb) & (\%) & \multicolumn{4}{c}{} &  & (\%) & (pb) \\
\tableline
50      & 100   & \multicolumn{4}{c}{-- (Relaxed \met\ threshold) --} & 43.0 & 0.8 & ${}^{+}_{-}$ &  ${}^{8.5}_{8.2}$     & 49 & 29.5 & --- \\
\tableline\tableline
75      & 100 & (150, 80) & (243, 249) &  4.4 & $5.8 \pm 0.5 {}^{+1.7}_{-1.4}$ & 8.3 & 0.8 & ${}^{+}_{-}$ & ${}^{3.4}_{3.2}$  & 15 &  9.2 & 4.4\\ 
75      & 120 & (300, 50) & (172, 318) & 15.7 & $1.5 \pm 0.3 {}^{+0.3}_{-0.2}$ & 5.5 & 0.5 & ${}^{+}_{-}$ & ${}^{2.7}_{2.6}$  & 12 & 6.2 & 14.8\\ 
75      & 140 & (200, 80) & (246, 278) &  2.4 & $5.8 \pm 0.4 {}^{+1.0}_{-1.6}$ & 3.6 & 0.2 & $\pm$ & 2.1           & 11 &  2.0 & 5.1\\
75      & 150 & (250, 60) & (198, 286) &  7.1 & $3.1 \pm 0.3 {}^{+0.4}_{-0.9}$ & 3.0 & 0.1 & $\pm$ & 1.9           &  8 &  6.1 & 8.1\\
75      & 160 & (300, 70) & (228, 339) &  2.0 & $4.2 \pm 0.4 {}^{+0.7}_{-0.8}$ & 2.6 & 0.1 & ${}^{+}_{-}$ & ${}^{1.8}_{1.7}$  &  6 &  12.9 & 3.3\\
90      & 100 & (100, 100) & (290, 266) &  1.8 & $7.7 \pm 0.5 {}^{+1.4}_{-1.5}$ & 6.0 & 0.7 & ${}^{+}_{-}$ & ${}^{2.7}_{2.5}$  &  8 &  31.8 & 1.7\\
100     & 100 & (0, 100) & (288, 250) &  2.8 & $4.9 \pm 0.4 {}^{+1.0}_{-1.1}$ & 4.6 & 0.7 & ${}^{+}_{-}$ & ${}^{2.2}_{2.0}$  &  7 &  25.4 & 2.7\\
100     & 150 & (200, 110) & (322, 330) &  0.3 & $9.2 \pm 0.5 {}^{+0.6}_{-1.3}$ & 1.3 & 0.1 & $\pm$ & 1.2           &  3 &  24.4 & 0.9
\end{tabular}
\end{table}

The largest noninstrumental background arises from the production of
\ttbar\ pairs in which one $t$ quark decays into jets and the other
decays into $b\ell\nu$, where $\ell$ = $e$, $\mu$, or $\tau$, and the
lepton is not detected. We generated 
$\ttbar \rightarrow b\ell\nu$ + jets events using the {\sc herwig} Monte
Carlo~\cite{herwig}. These were subjected to a detailed detector
simulation based on {\sc geant}~\cite{geant} and reconstructed with
the same reconstruction program used for data. We assumed the
\ttbar\ production cross section of $5.9\pm1.6$~pb~\cite{top-xsect},
which yielded a prediction of
$3.1\pm0.2$(stat)${}^{+1.4}_{-1.3}$(syst) background events. 

Comparable backgrounds come from the production of $W$ and $Z$ bosons.
Substantial \met\ can arise in events with a $W$ boson decaying to leptons where 
the charged lepton is not identified, and in events with
$Z \rightarrow \nu \nu$ or $Z \rightarrow \tau \tau$ decays. To
estimate these backgrounds, we generated Monte Carlo samples for $W$
boson events with {\sc vecbos}~\cite{vecbos} (quark hadronization
simulated using {\sc isajet}~\cite{isajet}), $Z$ bosons with {\sc
pythia}~\cite{pythia}, and $WW$ and $WZ$ events with {\sc isajet}. The
detector response was modeled as for the
\ttbar\ sample. From all vector boson production sources, we predict
$2.8 \pm 0.8 {}^{+0.7}_{-0.5}$~events, 85\% 
of which are from $W\rightarrow \ell\nu$ and $Z \rightarrow \nu \nu$ decays.

\begin{figure}[h]
\centerline{\psfig{figure=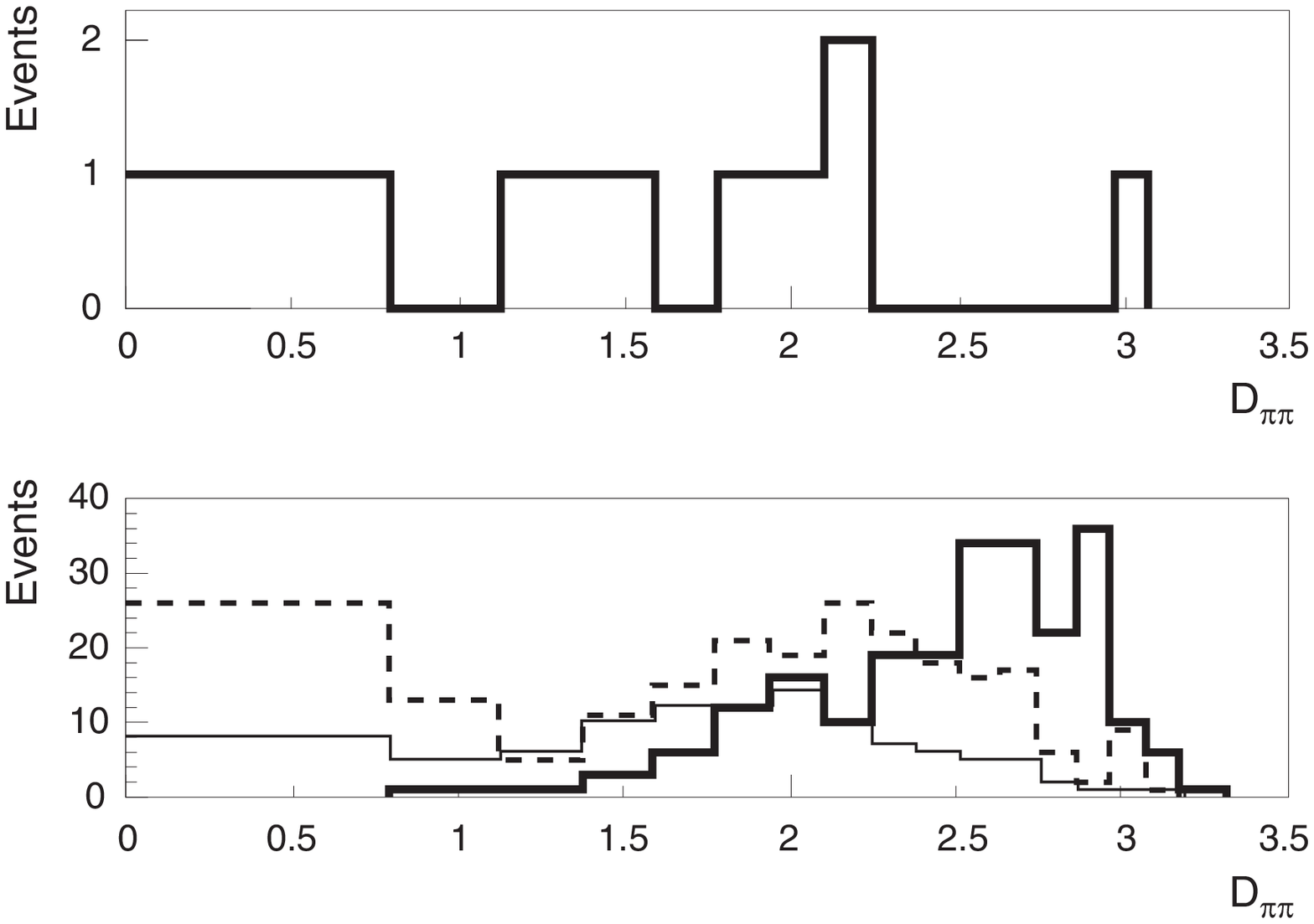,width=4.0in}}
\caption{
Sample \dpipi\ distributions used in the Bayesian shape fitter. Note
that the bins do not have uniform widths. The top plot shows data passing
the analysis requirements with \met~$>$ 75~GeV and \Ht~$>$ 150~GeV
(eight candidate events are accepted). The lower plot shows the
\dpipi\ distributions for QCD multijet (thick line), \ttbar\ (dashed line), and
an MLES sample with $\mzero = 250$~\gevcc\ and $m_{1/2} = 60$~\gevcc
(thin line) for events passing the same requirements. The
normalizations for the QCD multijet, \ttbar\, and MLES plots are 55.9, 7000,
and 350~\ipb, respectively.  }
\label{figure:dpipi}
\end{figure}

A major source of background is from events with three or more jets, where 
one (or more) is mismeasured, yielding apparent \met. To
determine this background, we used events from $56$~\ipb\
of data collected with a trigger requiring at least one jet with 
$E_T>85$~GeV. The trigger was fully efficient
for events containing a jet with $E_T > 115$~GeV. Events with
$\met < 50$~GeV were used to determine the instrumental background to events
with larger
\met\ using two different estimations. The primary method relied on a
Bayesian shape analysis~\cite{hist-fitter}. We define the quantity
$\dpipi = \sqrt{(\delta \phi_1 - \pi)^2+(\delta \phi_2 - \pi)^2}$, which has a distribution
that is strongly peaked at large \dpipi\ for events with apparent \met\
due to mismeasured jets.  For \ttbar\ and signal the
distribution is less peaked, as shown in Fig.~\ref{figure:dpipi}. For
the multijet events described previously, the shape of this
distribution is found to be nearly independent of the \met\
threshold. To determine the multijet contribution, we performed a
three-component (\ttbar, multijet, and signal) fit to the shape of the
\dpipi\ distribution in the data. The backgrounds quoted in
Table~\ref{table:optimization} include the multijet contribution, as
determined in this fit. As a check, we fit the \met\ spectrum of our
event sample between 25 and 50~GeV to an exponential in \met; extrapolation to
higher \met\ yielded a prediction in agreement with the fit to \dpipi,
as shown in Table~\ref{table:qcdfit}.

\begin{table}[h]
\caption{Comparison of the number of background events expected from QCD multijet
sources, as obtained from fits to \dpipi\ and from extrapolations from
lower \met\ (see text). {\em The uncertainties in the extrapolation do not include
the systematic uncertainty due to the dependence on the choice of functional form.}}
\label{table:qcdfit}
\begin{tabular}{c c c c}
$\met_{\rm thresh}$ & ${H_T}_{\rm thresh}$  & Bayesian Fit to \dpipi\ &
Extrapolation \\ (GeV) & (GeV) & & \\
\tableline
75 & 100 & $2.5 \pm 2.6$ & $2.8 \pm 0.9$ \\
75 & 150 & $0.8 \pm 1.6$ & $1.7 \pm 0.3$ \\
100 & 100 & $0.7 \pm 1.6$ & $0.6 \pm 0.1$
\end{tabular}
\end{table}

To check these background calculations, we relaxed the \met\ threshold
to 50~GeV, and obtained predictions of $7.6 \pm 0.8
{}^{+2.9}_{-2.1}$~events from \ttbar\ and $W$ and $Z$ boson
production, and $35.4 \pm 7.9$~events from QCD multijet, for a total of $43.0
\pm 0.8 {}^{+8.5}_{-8.2}$~events from background, as shown in 
Table~\ref{table:optimization}. We observed 49~events in the data.

We note that for all the entries in Table~\ref{table:optimization} the
number of observed events is greater than the number predicted from
background. The results are highly correlated, since most rows are
subsets of previous rows. The probability for obtaining at least the
number of events observed for any of the listed cutoffs is more than
2\%, and we therefore interpret our result as a constraint on the
\mzero\ and \mhalf\ parameters of MLES. Simulating squark and gluino production
and decay with {\sc isajet}, followed by the same detector response
and event reconstruction as in our previous simulations, we generated
samples at several values of \mzero\ and \mhalf, all with the MLES
parameters $\tanb = 2$, $A_0 = 0$, and $\mu < 0$. Using the
next-to-leading order squark and gluino production cross sections from
{\sc prospino}~\cite{prospino}, and a Bayesian technique with a flat
prior for the signal, we determined 95\% confidence level limits on
the parameters.

\begin{figure}[h]
\centerline{\psfig{figure=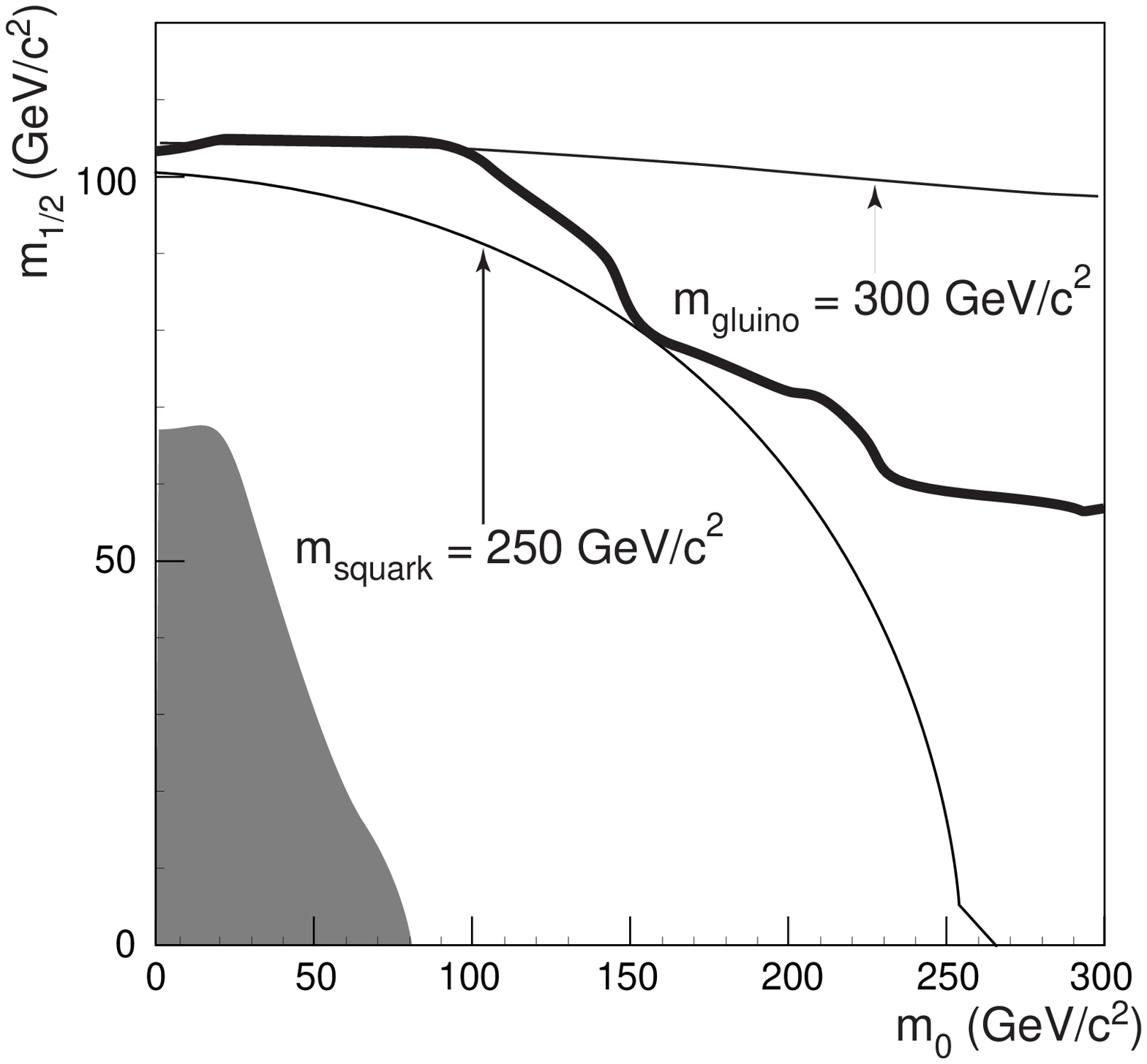,width=4.0in}}
\caption{
The exclusion contour obtained in this analysis (heavy line), the
region below which is excluded at the 95\% confidence level. The
thin lines are contours of constant squark or gluino mass in the
\mzero\ -- \mhalf\ plane, as indicated. In the shaded region MLES does
not contain electroweak symmetry breaking, and is excluded apriori.}
\label{figure:limit}
\end{figure}

\begin{figure}[h]
\centerline{\psfig{figure=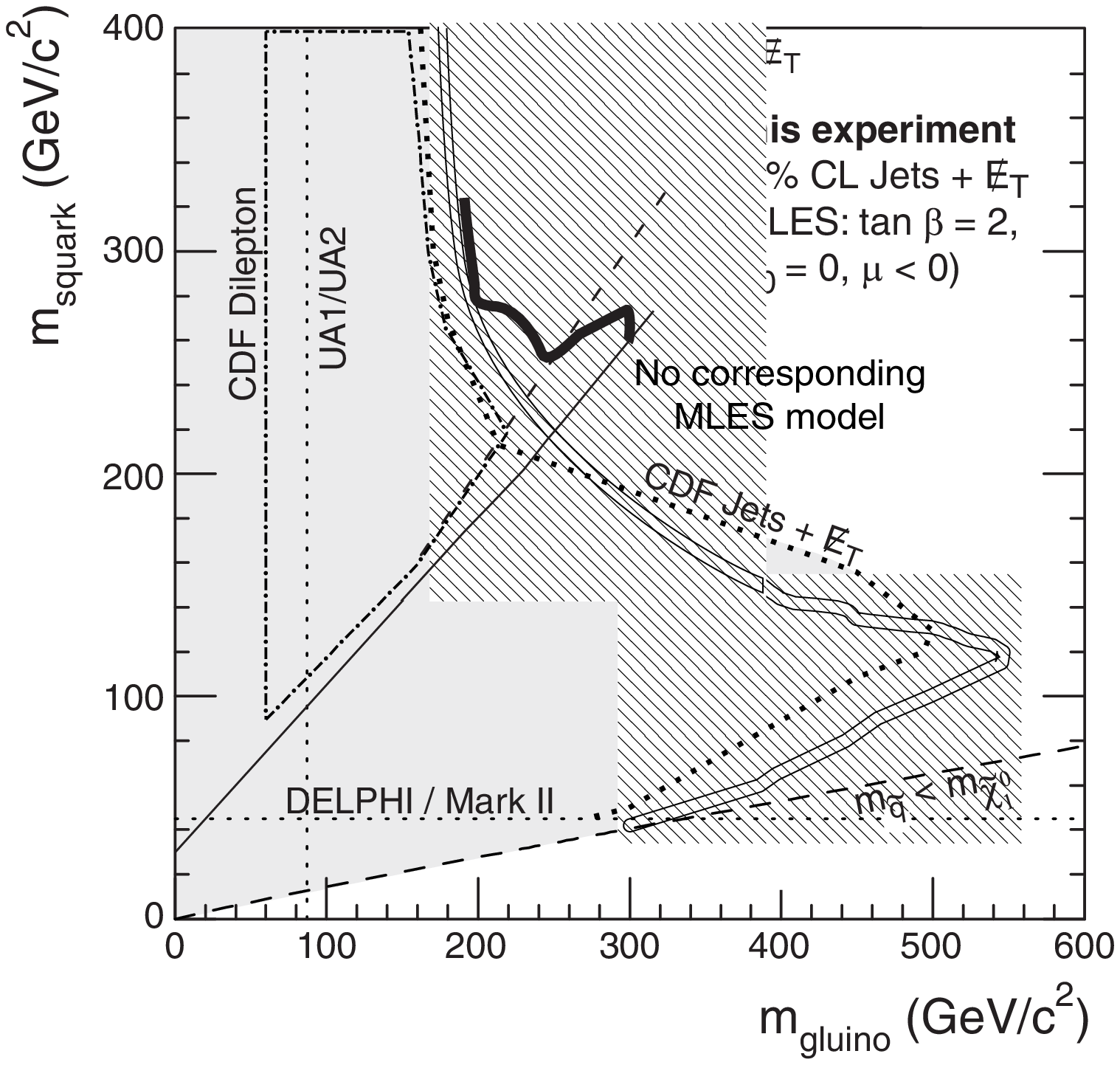,width=4.0in}}
\caption{The limit from this analysis in the $(\mgluino,\msquark)$
mass plane (``This experiment'').  The figure
also shows curves of previous limits from
jets and \met\  from \D0~\protect\cite{d0:sqgl,claes} (hatched)
using 7.2~\ipb\ of data and MSSM parameters 
$\tanb = 2$ and $\mu = -250$~\gevcc, the jets and \met\ limit
from CDF \protect\cite{cdf:sqgl} based on 19~\ipb\ of data with 
$\tanb = 4$ and $\mu = -400$~\gevcc\ (thick dots), the dilepton CDF
limit~\protect\cite{lala2} from 19~\ipb\ of data with MLES parameters
$\tanb = 4$ and $\mu <0$ (dashed-dotted),  and
limits using only direct decays from UA1/UA2
\protect\cite{ua} (dotted) and DELPHI/Mark II\protect\cite{delphi} (dotted). 
More recent model dependent CERN LEP limits are given in the text. All
limits are at the 95\% confidence level.  The region below the
diagonal dashed line is excluded because there the squark is the LSP.}
\label{figure:mgmq}
\end{figure}

Figure~\ref{figure:limit} shows the region excluded by this analysis.
Excluded are all MLES models with $\msquark < 250$~\gevcc.
For small \mzero, we exclude $\mgluino < 300$~\gevcc, and for
$\msquark = \mgluino$, we exclude masses less than 260~\gevcc. 
In Fig.~\ref{figure:mgmq}, we show the exclusion contour in the $(\mgluino,
\msquark)$ plane and compare our results to those from other
experiments. We extend significantly the limits on squarks
and gluinos, especially in the region where 
$\mgluino > \msquark$.  Within MLES models,  for negative
$\mu$ and $\tanb = 2$, the CERN LEP limits on charginos
$\mchargino <86$ to 45~\gevcc \cite{pdg} translate roughly to a limit
on \mhalf\ of 45~\gevcc\ for small \mzero, and 86~\gevcc\ for large
\mzero.  Our limit on \mhalf\ ranges between 100~\gevcc\ for small
\mzero, and 60~\gevcc\ for large \mzero. There are no self-consistent
MLES models below the solid diagonal line.

In summary, we have searched for events with large \met\ and multiple
jets, and observe no statistically significant excess of events beyond
expectations from SM processes.  This null result is
interpreted in the context of MLES as an excluded region in the
$(\mzero, \mhalf)$ plane.

We thank the Fermilab and collaborating institution staffs for
contributions to this work and acknowledge support from the 
Department of Energy and National Science Foundation (USA),  
Commissariat  \` a L'Energie Atomique (France), 
Ministry for Science and Technology and Ministry for Atomic 
   Energy (Russia),
CAPES and CNPq (Brazil),
Departments of Atomic Energy and Science and Education (India),
Colciencias (Colombia),
CONACyT (Mexico),
Ministry of Education and KOSEF (Korea),
and CONICET and UBACyT (Argentina).

\end{document}